# THE NEW DYNAMICS OF OPEN SOURCE: RELICENSING, FORKS, AND COMMUNITY IMPACT[1]

**Dawn Foster**[2]
CHAOSS
London, UK
dawn@dawnfoster.com

**ABSTRACT**

Many popular open source projects are owned and driven by vendors, and in today's difficult economic climate, those vendors are under increasing pressure from investors to deliver a strong return on their investments. One response to this pressure has been the relicensing of popular open source projects to more restrictive licenses in the hopes of generating more revenue, disrupting the idea of open source as a digital commons. In some cases, relicensing has resulted in a hard fork of the original project. These relicensing events and resulting forks can be disruptive to the organizations and individuals using these open source projects. This research compares and contrasts organizational affiliation data from three case studies based on license changes that resulted in forks: Elasticsearch / OpenSearch, Redis / Valkey, and Terraform / OpenTofu. The research indicates that the forks resulting from these relicensing events have more organizational diversity than the original projects, especially when the forks are created under a neutral foundation, like the Linux Foundation, rather than by a single company.

*Keywords:* Open source software, fork, licensing, organizational affiliation, CHAOSS

## 1    Introduction

This paper seeks to understand how the organizational dynamics of an open source project evolve following a relicensing event, especially when that relicensing event results in a fork of the original project. This topic is timely because there has been a growing trend of open source projects that have recently been relicensed [1], which has resulted in new forks of those projects that have been gaining traction [2]. These relicensing events and resulting forks can be disruptive to the organizations and individuals using these open source projects.

[1] Paper and data available from: https://github.com/chaoss/wg-data-science/tree/main/publications
Citation: Foster, D. (2024, November 13-14). The New Dynamics of Open Source: Relicensing, Forks, and Community Impact. OpenForum Academy Symposium 2024, Boston, Massachusetts.
[2] The author's work and the CHAOSS Data Science Initiative are funded by the Alfred P. Sloan Foundation.

Many popular open source projects are owned and driven by vendors,[3] and in today's difficult economic climate, those vendors are under increasing pressure from investors to deliver a strong return on their investments [3]. One response to this pressure has been the relicensing of popular open source projects to more restrictive licenses in the hopes of generating more revenue for the vendor that is driving that open source project [3], thus disrupting the idea of open source as a digital commons. Vendors that have relicensed their open source projects recently have included Elastic, HashiCorp, MongoDB, and Redis.[4] These relicensing events are disruptive to the organizations and individuals using these open source projects who must decide (often on short notice) whether they can keep using the project under the terms of a new license, and if not, they need to assess the effort required to migrate to another technology.

Some, but not all, cases of relicensing have resulted in a hard fork of the original project, which may provide an option for those organizations and individuals who decide that they cannot or will not continue using the original project under the terms of a new license. These hard forks, also referred to as hostile forks or variant forks, are created with the purpose of starting a new branch of development that diverges from and often competes with the original project under a new governance model for the fork [4]. Hard forks are in contrast to the social forks popularized by GitHub, which are often used for collaboration while contributing to the original project using a pull request model [5]. This research is focused on hard forks, so throughout this paper, all references to forks are hard forks unless otherwise specified.

Profitability, relicensing, and forking are all related, and under a certain set of circumstances, these three things can become part of a chain of events that leads from one to the next: profitability concerns can lead to relicensing, which can lead to forking. First, profitability concerns: when a vendor's revenue streams rely solely or mostly on an open source project, their investors and shareholders can become concerned that the vendor is not making as much profit as possible from the project. This is compounded if other companies (often large cloud providers) are competing for the same customers and profiting from the open source project that the vendor has put substantial resources into developing. Second, relicensing: these concerns about profitability lead to pressure from investors to put the open source project under a new license with more restrictions than would be possible under an open source license. These restrictions often make it more difficult for cloud providers or other companies to profit from the newly relicensed open source project. This relicensing event is only possible when the vendor controls the governance of the project, thus allowing the vendor to make unilateral decisions on behalf of the project, and in certain legal situations (e.g. contributor license agreement with copyright assignment) that allow for the relicensing event. Third, forking: after the vendor has relicensed the project from an open source license to a non-open source license, the users and contributors become concerned that they can no longer use the project in the ways that they were allowed to under the freedoms of the prior open source license. If there is enough support and resources to commit to forking the project and maintaining that fork over time, then a new fork may emerge. Because relicensing can only happen under certain governance structures and legal situations, these new forks are unlikely to be relicensed in the future, which might make them a safer choice for users. Any of these three events can also happen in isolation, but this research is focused on this specific chain of events where profitability concerns lead to relicensing, ultimately leading to a fork of the project.

---

[3] This paper uses the term "vendor" deliberately to refer to independent software vendors (https://en.wikipedia.org/wiki/Independent_software_vendor) whose primary business is to sell software / services, so "vendor" is a subset of the term "company". Under this definition and for this paper, Elastic, Redis, and HashiCorp are all vendors who sell software / services based on the open source projects that they have recently relicensed, while others who use and contribute to projects (e.g., Amazon, Google, Microsoft) are referred to as "companies".

[4] https://github.com/chaoss/wg-data-science/blob/main/dataset/license-changes/license_changes.csv

This chain of events from profitability concerns to relicensing to forking creates turmoil for the users of a project. However, there is very little academic research on this topic, which is what motivates this research on forks that result from relicensing. This research compares and contrasts data from three case studies based on license changes that resulted in forks: Elasticsearch with fork OpenSearch, Terraform with fork OpenTofu, and Redis with fork Valkey. This research uses a data-driven approach to answer the following research question: ***How do the organizational dynamics of an open source project evolve following a relicensing event, both within the original project and its resulting fork?***

The remainder of the paper is organized as follows. Section 2 contains a review of the literature. The methodology is presented in Section 3. Section 4 contains the results from the three case studies. A discussion of the results along with implications and limitations can be found in Section 5. Finally, Section 6 contains the conclusion to the paper.

## 2    Literature Review

While there is literature about vendor profitability, relicensing, and forks, no research could be found that focused on all three elements in this chain of events from profitability concerns to a vendor relicensing decision that ultimately results in a fork. This is likely because it is a rare, and recent phenomenon with few projects relicensed[5] by vendors to non-open source licenses and even fewer cases where the vendor relicensing resulted in a fork.[6] Because the cases in this study are limited to ones that resulted in a fork, the literature on forking will be covered first followed by relicensing and finally a discussion of vendor profitability as it relates to relicensing and forking.

### 2.1    Forking

The early open source licenses (e.g., MIT and GPL) started being used in the late 1980s, and while the term open source wouldn't be coined until much later, this was effectively the beginning of open source licensing as we know it today [6].  With the creation of open source licenses, came the forking of open source projects. A study by Robles and González-Barahona (2012) found only 13 forks prior to 1998. They went on to say that while forking is a freedom associated with open source software, it is considered a disruptive event that should be avoided whenever possible [7]. While hard forks cannot always be avoided, they are still a relatively rare phenomenon despite tools like GitHub lowering the bar and making it easier for people to create both social forks and hard forks [4]. Wheeler (2015) describes why hard forks are a disruptive and rare phenomenon [9]:

> *"Creating a fork is a major and emotional event in the FLOSS community. It [is] similar to a call for a "vote of no confidence" in a parliament, or a call for a labor strike in a labor dispute. Those creating the fork are essentially stating that they believe the project's current leadership is ineffective, and are asking developers to vote against the project leadership by abandoning the original project and switching to their fork. Those who are creating the fork must argue why other developers should support their fork; common reasons given include a belief that changes are not being accepted fast enough, that changes are happening too quickly for users to absorb them, that the project governance is too closed to outsiders, that the licensing approach is hampering development, or that the project's technical direction is fundamentally incorrect."* (Wheeler, 2015)

---

[5] https://github.com/chaoss/wg-data-science/blob/main/dataset/license-changes/license_changes.csv
[6] These are described in more detail in section 3.1 where the selection of cases is discussed.

Forking is a disruptive event, and forks usually evolve in parallel to the original project with no guarantee that either of them will ultimately be successful [7]. Businge et. al., (2022) found that hard forks are created and maintained by different people from the ones that stay to maintain the original project and that there is usually little to no interaction with the original project as the projects diverge and evolve in different directions[4].

### 2.2    Relicensing dynamics relative to forking

Next, the literature about relicensing comes into the discussion with a focus on the literature about the dynamics between relicensing and forking. Very few forks are a result of legal issues or relicensing [4], and there has been little academic research that is focused specifically on vendors putting their open source projects under non-open source licenses as the impetus for the creation of hard forks. This phenomenon results in significant disruption where the impacts are felt not just by the open source projects themselves, but also by companies using the open source software in their own infrastructure or incorporating it into their projects.

One of, if not the first, examples of an open source project fork resulting from a license change is the X.org fork created in 2004 from XFree86 when the creator of the project moved to a license that was not GPL-compatible, which ultimately resulted in XFree86 being abandoned after X.org was endorsed by the companies and developers using and contributing to the project [9]. There have also been many examples over the years of projects moving from one open source license to a more restrictive open source license for a variety of reasons, including a desire for greater revenue growth [10].

## 2.3    Vendor profitability as related to relicensing and forking

Recently, this pattern of relicensing to increase profitability seems to be happening more frequently with about 18 projects relicensed by vendors since 2018[7] from open source licenses to licenses that are not open source licenses as defined by the Open Source Initiative (OSI).[8] In many cases, these vendors are continuing to make the source code available and positioning the projects as "open" after the relicensing event. This technique is known as "open washing" and can mislead people into thinking that a project is still open source when it is not [2]. This appears to point to a tension where companies want the good image of being open source but cannot commit to preserving the freedoms of open source, which is a break with the open source ethos.

The increasing involvement of vendors as open source becomes more professional may be changing how we think about forking[4]. Vendor ownership of open source projects creates a power imbalance where that vendor has control over the governance of the project and can make unilateral decisions, which can increase the risk of a license change [8]. When vendors lead open source projects, there is a risk of organizational and strategic conflict that creates tension between the goals of the vendor and the other users and contributors to the open source project, which can result in a fork [7]. Riehle (2023) discussed how vendors often want to prevent competitors from using their open source software to compete with them [3]:

> *"Vendors license away from open source to source available typically only if they feel they need the goodwill of open source less urgently than before. As a business, these vendors' products probably matured and already reached the channel–product fit. … Once a product matures, for example, by reaching the channel–product fit, the open source strategy will lose some significance, and the need to increase profitability and return on investment for the venture capitalists behind the vendors will take over."* (Riehle, 2023)

Interestingly, relicensing appears to not have the desired financial benefits. Stephens (2024) from the RedMonk analyst firm took an in-depth look at several vendors (including a few of the case studies in this paper) that have relicensed their open source projects and found no evidence that relicensing has an impact on the vendors' financial results, despite this being the one of the primary reasons for the decision to move to a non-open source license [1]. In the case of Elastic, they've recently backtracked on their decision and have added an open source license (AGPLv3) on August 29, 2024 in addition to the non-open source licenses that they moved to in 2021 [11].

---

[7] https://github.com/chaoss/wg-data-science/blob/main/dataset/license-changes/license_changes.csv
[8] https://opensource.org/licenses

To understand the tension between preserving the freedoms of open source in a project and the commercial interests of a single company, we need to understand the relationship between the open source project and the company. The relationship between companies and open source projects is mutually beneficial and important because company participation influences the sustainability of open source projects [12]. Organizational participation from companies can help increase the survival and health of open source projects, especially when companies are collaborating within open source foundations [13]. On the other hand, having a single dominant company can put projects at risk of failure. Companies often make business decisions that can result in no longer funding employees to work on an open source project, which can result in projects losing a critical number of maintainers at one time [14]. Therefore, it's important to understand the organizational dynamics within open source projects and the impacts they have on the open source projects that we all rely on and use. Vendors are driving many open source projects, and those vendors are being increasingly pressured by investors and shareholders to improve profitability. Relicensing events and subsequent forks occur under this pressure, even though it is not clear that relicensing improves profitability or other financial outcomes [1].

## 3  Methods

### 3.1  Case studies

A case study approach [15] is used for this research to get insight into the following research question: *How do the organizational dynamics of an open source project evolve following a relicensing event, both within the original project and its resulting fork?* To get insight into this question, this research compares and contrasts data from three case studies of projects undergoing license changes that resulted in forks: Elasticsearch with fork OpenSearch, Redis with fork Valkey, and Terraform with fork OpenTofu. These case studies represent three different scenarios where the details and data can be found in the results section of the paper:

1. Almost all contributions to the original project came from employees of the original vendor and the fork was created by new contributors and owned by a single company (Elasticsearch / OpenSearch).
2. Almost all contributions to the original project came from employees of the original vendor and the fork was created by new contributors as a foundation project (Terraform / OpenTofu).
3. The original project had significant contributors who were not employed by the original vendor and the fork was created by those existing contributors as a foundation project (Redis / Valkey).

Open source projects that move from an open source license to a non-open source license and also result in a fork is a rare event. These vendor relicensing events have become more common since 2018 when MongoDB relicensed their open source project followed by several others in the last few years [16]. Out of these recent relicensing events, there were only two more hard forks found that were sustained, and both of them were a result of the same licensing changes as the case studies selected for this research. Elastic also relicensed Kibana, which resulted in an Amazon fork called OpenSearch Dashboards, which was recently put under the Linux Foundation. Hashicorp relicensed Vault at the same time as Terraform, and Vault was forked into a new project under the Linux Foundation called OpenBao. Since these two additional forks were generated by the same relicensing events, only the most popular projects were selected: Terraform and Valkey both have more stars and forks than Vault and OpenBao, and Elasticsearch and OpenSearch have more stars and forks than Kibana and OpenSearch Dashboards.

### 3.2     Data collection and Analysis

While there are many valuable ways to contribute to open source projects, the decision was made to focus on commits for this first paper. Commits provide a way to focus on who is doing the work of developing the code used in the project, and commit data can be consistently used for all of the open source projects in these cases. While commits are not perfect, and they leave out some essential work and other types of contributions (e.g., planning, architecture, community), commits are good enough to get an understanding of the organizational dynamics in each project to answer the research question, but they do not represent every contribution to a project.

Sometimes numbers of commits alone aren't a good indicator of the amount of work, so most of the results also look at the numbers of lines added and lines deleted to better understand the magnitude of the contributions.  Both types of data points can be misleading in isolation. Commits can be misleading because they may include a large number of commits with very small changes or a single commit that may have a large number of lines of code changed. Similarly, lines of code added and deleted can be inflated in certain types of changes, like refactoring or formatting updates that impact a large number of lines of code with little overall impact to the functionality of the project. Evaluating commits together with lines of code added and deleted can provide a more complete picture.

The biggest challenge with identifying the organizational dynamics in open source projects is that the organizational affiliation data is rarely accurate enough to use without doing an often significant amount of manual cleanup.[9] Job changes also need to be considered because it is possible that an individual works for one employer for part of a time period and another employer for a different time period; however, there were no cases of this found in any of the projects in any of the time periods. In several cases where an individual moved to a new employer, they stopped contributing to the project.

The commit data along with data about the author of the commit was gathered from the GitHub GraphQL API for the specified time period and repository using a custom Python script.[10] Some of the organizational affiliation data was obtained in an automated fashion for individual contributors who have populated the "Company" field in their GitHub profile or made commits from a company email address. Some additional cleanup on the author's "Company" field was required to take free form text (e.g., @elastic, Elastic, Elastic Inc) and output a consistent naming format for each company. This was done using simple pattern matching on the "Company" field. To determine organizational affiliation from email address data, the scripts looked for domain names that matched to email domains from companies known to contribute to the project and ignored common personal domains (e.g., gmail.com, outlook.com), so even if a single commit was from a company email address, an individual could be matched to that organization. The manual data cleanup described below is used if neither of these methods made an automated match.

---

[9] https://chaoss.community/practitioner-guide-organizational-participation/

[10] https://github.com/chaoss/wg-data-science/blob/main/dataset/license-changes/fork-case-study/commits_people.py

The manual data cleanup for these case studies came from a variety of sources, including LinkedIn, online searches, and talking to people involved in the projects to figure out where people work. Because this work is highly manual and time consuming, it isn't practical to do this for every person when you're looking at large, open source projects like the ones in these case studies. To make it a more manageable task, the numbers of commits were used as a starting point to only manually search for employer data from people who had made 10+ commits for larger projects and 5+ commits for smaller projects. Details about the data for each project along with the code used, comments about where manual data was found, and other data cleanup tasks can be found in a release for replication, validation, and inspection.[11]

## 4    Results

### 4.1    Elasticsearch

The Elasticsearch project is owned by a vendor commonly referred to as Elastic (legal name is Elasticsearch BV), a publicly traded company founded in 2012.[12] Elasticsearch was an open source project under the Apache 2.0 license until February 3, 2021 when the project was relicensed under the Server Side Public License (SSPL) and the Elastic License[13] - neither of which are open source licenses under the OSI definition. However, the way this was done was a bit misleading with the blog post announcing the license change titled, "Doubling Down on Open", which can be described as "open washing" as discussed earlier in the Literature Review section. The primary reason stated for the relicensing was due to a disagreement with Amazon and a concern that Amazon was profiting from Elasticsearch without the collaboration expected from open source norms of behavior.[14] However, it's commonly believed that pressure from investors is driving relicensing decisions, like this one [1].

On August 29, 2024, Elastic announced that they were adding the AGPLv3 (an open source license) as an additional licensing option in addition to the other licenses, effectively making

---

[11] https://github.com/chaoss/wg-data-science/releases/tag/v1.0-OFA-2024
[12] https://www.crunchbase.com/organization/elasticsearch
[13] https://www.elastic.co/blog/licensing-change
[14] https://www.elastic.co/blog/why-license-change-aws

Elasticsearch open source again.[15] Because this is a very recent change, there is not enough data to compare whether the addition of an open source license changes the organizational affiliation data for contributors or whether additional non-Elastic employees will contribute in larger numbers. However, this date was used as the end date for one year of analysis, so that the Elasticsearch contribution data is only for the time period when they were not an open source project.

In Table 1, contributions to the Elasticsearch project can be found across three different time periods for people making 10 or more commits in the https://github.com/elastic/elasticsearch repository.
- 1 Year before the relicense (2020-02-03 - 2021-02-03): This is when Elasticsearch was an open source project under the Apache 2.0 license. This time period offers a baseline comparison with the time periods after the project was relicensed.
- 1 Year after the relicense (2021-02-03 - 2022-02-03): This is the first year that Elasticsearch was under a proprietary (non-open source license).
- 1 Year before adding the AGPLv3 (2023-08-29 - 2024-08-29): This is an additional time period to understand whether there have been any recent changes to the contribution data.

As you can see from the table, Elastic employees have consistently made over 95% of the lines added to and deleted from Elasticsearch with almost no participation from other contributors from outside of the vendor. In this case, the relicensing made very little difference to the contributor makeup of the project because Elastic never really built a strong contributor base outside of the people who were employed there. While there was little impact on the contributor base, there was an impact on the users / consumers of Elasticsearch who were put in the position of being forced to decide whether to continue using it, and if so, under which of the two available licenses.

---

[15] https://www.elastic.co/blog/elasticsearch-is-open-source-again

**Table 1: Elasticsearch[16] - Number of Additions and Deletions from people making 10 or more commits in https://github.com/elastic/elasticsearch**

| Timeframe | Org Affiliation | People | Commits | Additions | Deletions |
|---|---|---|---|---|---|
| 1 Year Before the Relicense (2020-02-03 - 2021-02-03) | Elastic employees with 10+ commits | 67 | 6,477 (92%) | 1,377,558 (96%) | 623,561 (97%) |
| | Non-Elastic employees with 10+ commits | 3 | 94 (1%) | 3,855 (<1%) | 740 (<1%) |
| 1 Year after the relicense (2021-02-03 - 2022-02-03) | Elastic employees with 10+ commits | 65 | 5,668 (91%) | 1,597,988 (96%) | 1,061,154 (98%) |
| | Non-Elastic employees with 10+ commits | 2 | 47 (1%) | 7,283 (<1%) | 2,178 (<1%) |
| 1 Year Before Adding AGPL (2023-08-29 - 2024-08-29) | Elastic employees with 10+ commits | 99 | 7,616 (95%) | 2,621,830 (95%) | 1,123,628 (97%) |
| | Non-Elastic employees with 10+ commits | 1 | 11 (<1%) | 326 (<1%) | 326 (<1%) |

Note: Percentages use the total number of additions / deletions, not just from people making 10 or more commits, which is why the numbers do not add up to 100%. The further away from totalling 100% indicates a larger long tail of contributions with more people making very small contributions.

### 4.2 OpenSearch

OpenSearch[17] was forked from Elasticsearch on April 12, 2021 by Amazon, a publicly traded company founded in 1994.[18] The OpenSearch fork was derived from Elasticsearch 7.10.2, which was under the Apache 2.0 license, and OpenSearch was also licensed under Apache 2.0, the same open source license as before.[19] The OpenSearch fork was similar to the original

---

[16] This data (including any updates) is being maintained and version controlled at https://github.com/chaoss/wg-data-science/blob/main/dataset/license-changes/fork-case-study/notebooks/elasticsearch.ipynb

[17] The history of OpenSearch is more complex than can be covered in the scope of this paper. This blog post has more details about Amazon's Open Distro for Elasticsearch, a predecessor to OpenSearch: https://aws.amazon.com/blogs/aws/new-open-distro-for-elasticsearch/

[18] https://www.crunchbase.com/organization/amazon

[19] https://aws.amazon.com/blogs/opensource/introducing-opensearch/

Elasticsearch project in that it was owned by a company (Amazon). Before the relicense, Amazon was selling Elasticsearch services to their Amazon AWS customers, and after the relicense, Amazon AWS created the OpenSearch fork so that it could continue to offer similar services to their customers. By making OpenSearch an open source project, it also allowed other people to use and contribute to the project. OpenSearch continued to be owned by Amazon until September 16, 2024 when they put the project under the Linux Foundation (LF), a neutral nonprofit; therefore, it ceased to be a vendor owned open source project.[20] Like with Elastic adding the AGPLv3 license, putting OpenSearch under the LF is a very recent change, and there is not enough data to compare the organizational affiliation data for contributors after that time, but it was used as the end date for one year of analysis so that recent contribution dynamics can be compared with earlier time periods.

With the dominance of Elastic employees driving Elasticsearch, Amazon was almost, but not quite, starting from scratch to create the OpenSearch fork. There were nine people who contributed to Elasticsearch in the year before the relicense who later contributed to OpenSearch. Five of these people were Amazon employees who were likely contributing to Elasticsearch to support Amazon's use of the project. One of them was a previous Elastic employee who left well before the relicensing event to join Amazon in November 2020, and another is someone who was making a fair number of contributions to Elasticsearch on behalf of Amazon before the relicensing event. Both of these people made considerable contributions to OpenSearch and were likely providing some continuity and knowledge of the Elasticsearch codebase, which would have been helpful when creating and maintaining the OpenSearch fork.

In Table 2, contributions to the OpenSearch project can be found across two different time periods for people making 10 or more commits in the https://github.com/opensearch-project/OpenSearch repository.
- 1 Year after the fork (2021-04-12 to 2022-04-12): This is one year of data shortly after the fork to use as a comparison with a more recent time period to understand how the project is evolving.
- 1 Year before moving under the LF (2023-09-16 to 2024-09-16): This year of data helps to understand how the project is evolving in ways that may have influenced the decision to move the project under the LF.

Like with Elasticsearch, most of the contributions to OpenSearch come from Amazon employees; however, to a lesser extent and with increases in organizational diversity over time. As you can see in Table 2, in the first year of the fork, a small number of Amazon employees (those with 10 or more commits) made 80% of total additions and 91% of total deletions to the code in the https://github.com/opensearch-project/OpenSearch repository. Only two people who didn't work for Amazon made 10 or more commits making up 7% of additions and 4% of deletions.

---

[20] https://www.linuxfoundation.org/press/linux-foundation-announces-opensearch-software-foundation-to-foster-open-collaboration-in-search-and-analytics

In the final year of the fork under ownership by Amazon before the project was moved under the Linux Foundation, the organizational diversity improved with only 63% of additions and 64% of deletions coming from Amazon employees making 10 or more commits. Six people who didn't work for Amazon made 10 or more commits making up 11% of additions and 13% of deletions, so participation from outside of Amazon is increasing. However, it's worth noting that one person who works at Aiven made most of the non-Amazon additions (31,718) and deletions (8175). It will be interesting to re-run this analysis six months to a year after OpenSearch has been under the LF.

**Table 2: OpenSearch[21] - Number of Additions and Deletions from people making 10 or more commits in https://github.com/opensearch-project/OpenSearch**

| Timeframe | Org Affiliation | People | Commits | Additions | Deletions |
|---|---|---|---|---|---|
| 1 Year After the Fork (2021-04-12 to 2022-04-12) | Amazon employees with 10+ commits | 7 | 246 (34%) | 296,720 (80%) | 224,179 (91%) |
| | Non-Amazon employees with 10+ commits | 2 | 110 (15%) | 26,995 (7%) | 10,799 (4%) |
| 1 Year before LF (2023-09-16 to 2024-09-16) | Amazon employees with 10+ commits | 40 | 923 (49%) | 237,781 (63%) | 48,894 (65%) |
| | Non-Amazon employees with 10+ commits | 6 | 242 (13%) | 42,863 (11%) | 9,936 (13%) |

Note: Percentages use the total number of additions / deletions, not just from people making 10 or more commits, which is why the numbers do not add up to 100%. The further away from totalling 100% indicates a larger long tail of contributions with more people making very small contributions.

### 4.3   Terraform

The Terraform project is owned by HashiCorp, a publicly traded company founded in 2012 focused on infrastructure cloud software.[22] They are in the process of being acquired by IBM with the transaction expected to close by the end of 2024.[23] Terraform was under the open source Mozilla Public License v2.0 (MPL 2.0) until August 10, 2023 when it was relicensed along with their other open source projects (e.g., Vagrant, Vault) to the Business Source License (BSL), which is not an open source license under the OSI definition.[24] The stated reasons for the relicensing event were similar to Elastic and Redis; the BSL license would put restrictions on commercial use and make it more difficult for other companies to compete with HashiCorp while not putting in the same amount of effort into the development of the project.

---
[21] This data (including any updates) is being maintained and version controlled at https://github.com/chaoss/wg-data-science/blob/main/dataset/license-changes/fork-case-study/notebooks/OpenSearch.ipynb
[22] https://www.crunchbase.com/organization/hashicorp
[23] https://www.globenewswire.com/news-release/2024/07/15/2913382/0/en/HashiCorp-Shareholders-Vote-to-Approve-Transaction-with-IBM.html#

In Table 3, contributions to Terraform can be found across two time periods for people making 5 or more commits to the https://github.com/hashicorp/terraform repository:
- 1 Year before the relicense (2022-08-10 to 2023-08-10): This is when Terraform was an open source project under the MPL 2.0 license. This time period offers a baseline comparison with the time periods after the project was relicensed.
- 1 Year after the relicense (2023-08-10 to 2024-08-10): This is the first year that Terraform was under a proprietary (non-open source license).

Similarly to Elasticsearch, Table 3 shows that Terraform had very few contributors who weren't employees of Hashicorp. In the year before the relicensing event (and in the year after) there were only 2 contributors to Terraform who were not affiliated with HashiCorp, and they both made a very small number of contributions. Since there were so few contributions from outside of the vendor, there was no substantial impact on the contributor community from the relicensing event, so the only people impacted would likely have been the users of Terraform, which can be seen in the data about the OpenTofu fork in the next section.

---

[24] https://www.hashicorp.com/blog/hashicorp-adopts-business-source-license

**Table 3: Terraform[25] - Number of Additions and Deletions from people making 5 or more commits in https://github.com/hashicorp/terraform**

| Timeframe | Org Affiliation | People | Commits | Additions | Deletions |
|---|---|---|---|---|---|
| 1 Year before relicense (2022-08-10 - 2023-08-10) | HashiCorp employees with 5+ commits | 21 | 971 (82%) | 202,612 (93%) | 81,019 (95%) |
| | Non-HashiCorp employees with 5+ commits | 2 | 13 (1%) | 84 (<1%) | 33 (<1%) |
| 1 Year after relicense (2023-08-10 to 2024-08-10) | HashiCorp employees with 5+ commits | 24 | 1,620 (91%) | 672,393 (90%) | 242,052 (93%) |
| | Non-HashiCorp employees with 5+ commits | 2 | 18 (1%) | 353 (<1%) | 354 (<1%) |

Note: Percentages use the total number of additions / deletions, not just from people making 5 or more commits, which is why the numbers do not add up to 100%. The further away from totalling 100% indicates a larger long tail of contributions with more people making very small contributions.

### 4.4　OpenTofu

OpenTofu was forked from Terraform on August 25, 2023 as a Linux Foundation project under the MPL 2.0.[26] There was an attempt before the fork to convince HashiCorp to rethink the decision and go back to an open source license with the OpenTofu Manifesto,[27] which had support from over 100 companies and 1000 individuals at the time of the fork.

In Table 4, contributions to OpenTofu can be found for people making five or more commits in the https://github.com/opentofu/opentofu repository for 1 year after the fork:
- 1 year after the fork (2023-09-05 to 2024-09-05): This is the first year after OpenTofu was forked from Terraform.

---

[25] This data (including any updates) is being maintained and version controlled at https://github.com/chaoss/wg-data-science/blob/main/dataset/license-changes/fork-case-study/notebooks/terraform.ipynb
[26] https://opentofu.org/blog/opentofu-announces-fork-of-terraform/
[27] https://opentofu.org/manifesto/

The fork was driven by a group of Terraform users from a variety of companies, but there were no contributors to OpenTofu who had previously contributed to Terraform, so they were starting from scratch with the codebase. While the manifesto had wider support, the actual contributions came from 31 people at 11 organizations who have made five or more contributions to the primary OpenTofu repository in the first year. The most substantial contributions have come from Spacelift, whose employees have made over half of the additions and deletions. Employees from Env0 and Scalr have also made quite a few of the contributions.

Table 4: OpenTofu[28] - Number of Additions and Deletions from organizations making 5 or more commits in https://github.com/opentofu/opentofu for 1 year after the fork (2023-09-05 - 2024-09-05)

| People | Organization | Commits | Additions | Deletions |
| --- | --- | --- | --- | --- |
| 10 | Spacelift | 328 | 88121 (55.21%) | 63992 (69.15%) |
| 6 | Env0 | 99 | 26507 (16.61%) | 12248 (13.23%) |
| 3 | Scalr | 47 | 12516 (7.84%) | 3374 (3.65%) |
| 3 | Harness | 17 | 2948 (1.85%) | 366 (0.40%) |
| 3 | Red Hat | 15 | 1605 (1.01%) | 159 (0.17%) |
| 1 | Hangzhou Dianzi University | 6 | 891 (0.56%) | 242 (0.26%) |
| 1 | Chainguard | 6 | 266 (0.17%) | 93 (0.10%) |
| 1 | lessops | 6 | 2017 (1.26%) | 226 (0.24%) |
| 1 | claranet | 6 | 118 (0.07%) | 20 (0.02%) |
| 1 | Cooby-inc | 5 | 72 (0.05%) | 69 (0.07%) |
| 1 | nvdnc | 5 | 68 (0.04%) | 11 (0.01%) |

Note: Percentages use the total number of additions / deletions, not just from organizations making 5 or more commits, which is why the numbers do not add up to 100%. The further away from totalling 100% indicates a larger long tail of contributions with more organizations making very small contributions.

### 4.5 Redis

The Redis project is owned by Redis (previously Redis Labs), a privately owned company founded in 2011 that provides software and services related to the project.[29] The Redis project was an open source project under the Berkeley Software Distribution 3-clause (BSD-3) until

---

[28] This data (including any updates) is being maintained and version controlled at https://github.com/chaoss/wg-data-science/blob/main/dataset/license-changes/fork-case-study/notebooks/opentofu.ipynb
[29] https://www.crunchbase.com/organization/redis

March 20, 2024 when the project was relicensed under the Redis Source Available License (RSALv2) and the SSPLv1[30] - neither of which are open source licenses under the OSI definition. This is contrary to the 2018 Redis blog post stating that the Redis open source project would always remain under the BSD license.[31] The Redis project has a complicated history that is outside of the scope of this research paper, but can be found in an LWN article written by Joe Brockmeier [17].

In the relicensing announcement on March 20, 2024, Redis made it clear that the reason for the new licenses was because of difficulties sustaining the company while cloud providers were profiting from the open source work and making it difficult for Redis to sustain their business with the project under an open source license. This is similar to the argument that Elastic made for their relicensing action.

In Table 5, contributions to the Redis project can be found across two different time periods for people making five or more commits in the https://github.com/redis/redis repository:
- 1 year before the relicense (2023-03-20 to 2024-03-20):  This is when Redis was an open source project under the BSD license. This time period offers a baseline comparison with the time periods after the project was relicensed.
- 6 Months after the relicense (2024-03-20 to 2024-09-20): Because the relicensing event was recent, we have less than a year of data to compare, so six months was the time period selected to make the comparison to one year of data easier. This was when Redis was under a proprietary (non-open source license).

Where the Redis project differs from Elasticsearch is in the number of contributions from people who were not employees of Redis. In the year leading up to the relicense, while Redis was still open source, there were substantial contributions from employees of other companies with twice as many non-employees making five or more commits. While Redis employees made more additions and deletions (which were inflated by a small number of very large change sets), employees of other companies made almost twice as many commits from about a dozen people who made substantial contributions.

In the six months after the relicense, all of the external contributors from companies like Amazon, Alibaba, Tencent, Huawei, and Ericsson who contributed over five commits to the Redis project in the year leading up to the relicense stopped contributing. Some made one to three commits shortly after the relicense, which likely indicated work already in progress. All of the contributors who made five or more commits after the relicense were Redis employees.

---

[30] https://redis.io/blog/redis-adopts-dual-source-available-licensing/
[31] https://redis.io/blog/redis-license-bsd-will-remain-bsd/

Table 5: Redis[32] - Number of Additions and Deletions from people making 5 or more commits in https://github.com/redis/redis

| Timeframe | Org Affiliation | People | Commits | Additions | Deletions |
| --- | --- | --- | --- | --- | --- |
| 1 year before relicense (2023-03-20 - 2024-03-20) | Redis employees with 5+ commits | 6 | 164 (28%) | 189,656 (80%) | 83,122 (74%) |
| | Non-Redis employees with 5+ commits | 12 | 319 (54%) | 28,334 (12%) | 16,684 (15%) |
| 6 Months after relicense (2024-03-20 - 2024-09-20) | Redis employees with 5+ commits | 7 | 154 (74%) | 38,270 (75%) | 10,464 (72%) |
| | Non-Redis employees with 5+ commits | 0 | 0 | 0 | 0 |

Note: Percentages use the total number of additions / deletions, not just from people making 5 or more commits, which is why the numbers do not add up to 100%. The further away from totalling 100% indicates a larger long tail of contributions with more people making very small contributions.

### 4.6   Valkey

Valkey was forked from Redis 7.2.4 on March 28, 2024 as a Linux Foundation project under the BSD-3 license driven by a group of people who had previously contributed to Redis with public support from their employers[33]. This is different from the OpenSearch fork because it was started as a foundation project by a group of companies, rather than by a single company, and because it was driven by people who had previously made substantial contributions to Redis.

In Table 6, contributions to Valkey can be found from people making five or more commits in the https://github.com/valkey-io/valkey repository in the first six months of the fork:
- First 6 months after the fork (2024-03-28 to 2024-09-28): Because the relicensing event was recent, we have less than a year of data to compare, so six months was the time period selected to make the comparison to Redis easier.

Although we only have data for the first six months of the project, this is enough to start to understand the organizational dynamics within the Valkey project. Valkey had 29 people employed at 10 companies in the first six months of the fork, and 18 of those people previously contributed to Redis and are now contributing to Valkey. There are a diverse set of contributors from a variety of companies with Amazon having the most contributors. The Google data is a bit misleading because while Google employees have made a large percentage of additions and deletions, most of those changes (~28K additions and ~30K deletions) can be attributed to two PRs, which were related to formatting changes. There are also many contributions coming from Ericsson, Huawei, and others. With Valkey as a Linux Foundation project from the beginning, it's not unusual to see contributions from a diverse set of organizations. Valkey has also been getting some attention recently due to the momentum experienced in quite a short period of time.[34]

---

[32] This data (including any updates) is being maintained and version controlled at https://github.com/chaoss/wg-data-science/blob/main/dataset/license-changes/fork-case-study/notebooks/redis.ipynb
[33] https://www.linuxfoundation.org/press/linux-foundation-launches-open-source-valkey-community

Table 6: Valkey[35] - Number of Additions and Deletions from organizations making 5 or more commits in https://github.com/valkey-io/valkey for the first 6 months after the fork (2024-03-28 - 2024-09-28)

| People | Organization | Commits | Additions | Deletions |
|---|---|---|---|---|
| 13 | Amazon | 149 | 18232 (18.31%) | 6288 (8.66%) |
| 1 | Tencent Cloud | 92 | 4859 (4.88%) | 2429 (3.35%) |
| 4 | Huawei | 76 | 3561 (3.58%) | 3016 (4.16%) |
| 2 | Ericsson | 45 | 5867 (5.89%) | 1954 (2.69%) |
| 2 | Google | 39 | 40698 (40.86%) | 38643 (53.24%) |
| 1 | Intel | 12 | 632 (0.63%) | 464 (0.64%) |
| 1 | Alibaba | 8 | 415 (0.42%) | 71 (0.10%) |
| 1 | @gnet-io | 8 | 104 (0.10%) | 73 (0.10%) |
| 2 | ByteDance | 7 | 3952 (3.97%) | 572 (0.79%) |
| 2 | Samsung | 5 | 48 (0.05%) | 48 (0.07%) |

Note: Percentages use the total number of additions / deletions, not just from organizations making 5 or more commits, which is why the numbers do not add up to 100%. The further away from totalling 100% indicates a larger long tail of contributions with more organizations making very small contributions.

---

[34] Stephens, R. (2024). Valkey Momentum: Seven Months In. RedMonk. https://redmonk.com/rstephens/2024/10/11/valkey-momentum/
[35] This data (including any updates) is being maintained and version controlled at https://github.com/chaoss/wg-data-science/blob/main/dataset/license-changes/fork-case-study/notebooks/valkey.ipynb

## 4.7 Summary

Table 7 provides a concise summary of the results from the six sections above aligned within the three scenarios described in the Methods section.

**Table 7: Summary of Results**

| **Scenario 1.** Almost all contributions to the original project came from employees of the original vendor and the fork was created by new contributors and owned by a single company. ||
|---|---|
| **Elasticsearch:** Contributors are mostly Elastic employees both before and after the relicense. | **OpenSearch:** Contributors are mostly from Amazon, but organizational diversity is gradually improving. |
| | |
| **Scenario 2.** Almost all contributions to the original project came from employees of the original vendor and the fork was created by new contributors as a foundation project. ||
| **Terraform:** Contributors are mostly HashiCorp employees both before and after the relicense. | **OpenTofu:** 31 people employed at 11 companies, but none previously contributed to Terraform. |
| | |
| **Scenario 3.** The original project had significant contributors who were not employed by the original vendor and the fork was created by those existing contributors as a foundation project. ||
| **Redis:** Strong organizational diversity before the relicense, but only Redis employees after. | **Valkey:** 29 people employed at 10 companies have contributed, and 18 of them moved from Redis. |

## 5 Discussion and Implications

This research shows the diversity of outcomes that can occur when a vendor relicenses an open source project that results in a fork of the original project. Looking at forks that were successful in the past (e.g., OpenOffice to LibreOffice, MySQL to MariaDB, OwnCloud to NextCloud), one thing they often had in common was that the forks were driven by people who were involved in the project [16]. These new forks that are resulting from vendor relicensed open source projects have very different organizational dynamics from the forks of the past. This research indicates that it is possible to have successful forks of a project even when there are few or even none of the contributors to the original project, as in the OpenSearch and OpenTofu cases.

While this paper only explores three cases with six total projects, it does shed some light on the research question: "*How do the organizational dynamics of an open source project evolve following a relicensing event, both within the original project and its resulting fork?*"

Starting with the original projects, Elasticsearch and Terraform had little to no change to the organizational dynamics within those projects. Employees of Elastic and HashiCorp were making well over 90% of the contributions in the year before the relicense and again in the year after with very few contributions from outside of those vendors, so the organizational dynamics were about the same before and after the relicense action. Redis on the other hand, shows a substantial difference in organizational dynamics, which is not unexpected given that there were employees from 12 other companies contributing to Redis who were disenfranchised by the licensing change. The employees from those 12 companies stopped contributing after the relicensing event, and most of them began contributing to the fork, instead, which brought the number of outside contributors with 5 or more commits in Redis to zero in the 6 months after the relicense.

The three forks also show some interesting differences. In the case of OpenSearch, which was owned by a company, in the first year of the fork (2021 - 2022) more of the contributions were coming from Amazon employees, and while this improved in the most recent year (2023 - 2024), Amazon employees were still making over 60% of the changes to the codebase with only six people who weren't Amazon employees making 10 or more commits. It will be interesting to see how the organizational dynamics evolve now that OpenSearch is a Linux Foundation project. The other two forks, Valkey and OpenTofu, were started not by a single company, but by a group of companies as projects under the Linux Foundation. Both of those forks had more organizational diversity than OpenSearch, and more organizational diversity than the original projects. For the people making 5 or more commits to the primary repository, Valkey had 29 people employed at 10 companies in the first six months of the fork, and OpenTofu had 31 people employed at 11 companies in the first year of the fork. Out of these three forks, the ones started under neutral foundations had more organizational diversity than the one started by a company.

## 5.1    Limitations and Future Research

This research covers only three case studies of open source projects that were relicensed and resulted in a fork, and it only assesses one aspect of those projects, organizational dynamics, so there are many opportunities for future research. The next step for this research project is to go beyond organizational dynamics to study other aspects of project health using additional CHAOSS metrics for these case study projects. For some projects, it might also be beneficial to perform this project health analysis beyond just the primary repository to look at other repositories where related work is happening.

It would be interesting to understand whether the recent changes to Elasticsearch (adding the AGPLv3 license) and OpenSearch (becoming a Linux Foundation project) impact the organizational affiliation data of these projects. For now, those changes are too recent to assess.

Expanding this research to additional cases would shed more light on whether the organizational dynamics we're seeing hold true for different types of projects in different situations. Right now, the dataset of projects that have relicensed and resulted in forks is quite small. It could be expanded to look at Elastic's Kibana project and OpenSearch Dashboards fork along with HashiCorp's Vault and fork OpenBao, but those might be very similar to the projects in these existing cases. If the relicensing of vendor open source projects continues, there might be more cases to study in the future.

In addition to the limited number of cases and single metric (organizational affiliation), the research is further limited because we could be working more closely with project community members who work within these projects. Best practice in open source research is to collaborate with the people who are involved in the projects being studied [18]. The author has had discussions about the data with experts on Elasticsearch, Redis, Valkey, and OpenTofu who already provided feedback and additional context, which was used to improve the data and results.

A limitation is the use of commit history as the only type of open source contributions. Commits are the essential type of contribution because all code contributions need to be made as a commit. However, open source projects require a lot more types of contributions to function. Especially in a user-oriented community, engagement could occur through other channels, for example in issues, forums, or instant messengers. Future research could expand the unit of analysis to include other types of contributions for a more comprehensive view of the open source project community.

## 5.2   Implications

While there is still more research to be done, there are already a number of implications for practitioners, researchers, and policy makers.

**Practice.** Vendors should carefully consider the decision to relicense an open source project. Would any additional financial benefits outweigh the downsides associated with a potential fork (e.g., loss of contributors, negative press)? There are also implications for organizations and individuals who are using or consuming open source projects driven by vendors who could potentially relicense that project in the future. When thinking about which open source projects to adopt, the organizational participation dynamics should be taken into account before committing the time and energy to adopt projects that are dominated by a single vendor.

**Research.** As discussed above in the limitations and future research section, there is more to study about vendor relicensing and the forks that result from those events. This paper also highlights the importance of looking carefully at organizational dynamics when performing research related to open source projects. An interesting question is to consider also the user base of the original and forked project, if such data could be obtained.

**Policy.** Like with practitioners, policy makers would be well served to consider the organizational dynamics in open source projects when making any policy decisions related to adoption of open source, since not all open source projects are created equally, and there is a real possibility that what was once an open source project may not be open source forever, and policies need to take these dynamics into account. Additionally, we've seen governmental organizations putting funding programs in place designed to improve the sustainability of open source projects over time. Those funding programs should consider the risk of funding projects that are controlled by a single vendor.

## 6 Conclusion

With the rise of cloud computing, large cloud companies (e.g., Amazon, Google) are creating commercial offerings based on open source projects owned by other companies, usually much smaller vendors. In many cases, these large cloud computing companies are not contributing back to these projects at a level commensurate with the benefit they are receiving from their use of the software. At the same time, vendors are under increasing pressure from investors to deliver a strong return on their investments. One way that vendors are choosing to do this is by relicensing popular open source projects in the hopes of generating more revenue, but this can also result in hard forks of those projects. These relicensing events and resulting forks can be disruptive to the organizations and individuals using these open source projects.

The forks that result from these relicensing events tend to have more organizational diversity than the original projects, especially when the forks are created under a neutral foundation, like the Linux Foundation, rather than being forked by a single company. It is still too early to understand the ultimate success or failure of these projects (both the original and the fork). These new forks have more organizational diversity, and projects with greater organizational diversity tend to be more sustainable [14]. However, it remains to be seen if these types of forks might be more sustainable than the original projects over time for any vendors who continue to struggle to meet the expectations of their investors.

## 7 Acknowledgements

The author's work and the CHAOSS Data Science Initiative are funded by the Alfred P. Sloan Foundation. Thank you to Amanda Brock and Stephen Walli for a variety of in depth conversations on this topic. Thank you to the following individuals for their feedback on portions of this paper and / or the data contained within it: Mirko Boehm, Cali Dolfi, Armstrong Foundjem, Matt Germonprez, James Humphries, Georg Link, Madelyn Olson, Cailean Osbourne, Alice Sowerby, Greg Sutcliffe, Jamie Tanna, Sophia Vargas, and Chan Voong. Thank you to members of the CHAOSS and OpenUK communities who have listened to me give talks on this topic and provided feedback or participated in discussions after those talks.